\documentclass[aps,pra,superscriptaddress, 10pt,twocolumn]{revtex4-2}
\usepackage{bm}
\usepackage{amsmath}
\usepackage{amssymb}
\usepackage{slashed}
\usepackage{float}
\allowdisplaybreaks
\usepackage{dcolumn}
\newcolumntype{d}[1]{D{.}{.}{#1}}
\newcolumntype{w}[1]{D{.}{.}{#1}}
\newcommand*{\centt}[1]{\multicolumn{1}{c}{#1}}

\usepackage{nicefrac}
\usepackage{physics}
\usepackage{graphicx}
\newcommand{\Za}{Z\alpha}
\newcommand{\vare}{\varepsilon}
\newcommand{\lbr}{\langle}
\newcommand{\rbr}{\rangle}

\begin{document}
\title{Two-body $\bm P$-state energies at $\bm{\alpha^6}$ order}

\author{Vojt\v{e}ch Patk\'o\v{s}}
\affiliation{Faculty of Mathematics and Physics, Charles University,  Ke Karlovu 3, 121 16 Prague
2, Czech Republic}

\author{Vladimir A. Yerokhin}
\affiliation{Max–Planck–Institut f\"ur Kernphysik, Saupfercheckweg 1, 69117 Heidelberg, Germany}

\author{Krzysztof Pachucki}
\affiliation{Faculty of Physics, University of Warsaw,
             Pasteura 5, 02-093 Warsaw, Poland}

\begin{abstract}
We present an analytical calculation of the complete $\alpha^6$ correction to energies of $nP$-levels of two-body systems
consisting of the spin-$0$ or $1/2$ extended-size particles
with arbitrary masses and magnetic moments.
The obtained results apply to a wide class of two-body systems  such as hydrogen, positronium,
muonium, and pionic or aniprotonic helium ion.
They are also relevant for light muonic atoms, whose accurate
theoretical predictions are required for extracting the nuclear charge radii.
\end{abstract}
\maketitle

\section{Introduction}
Two-body systems, such as hydrogen and hydrogen-like ions \cite{codata18}, muonic hydrogen \cite{pohl_10},
muonic helium ion \cite{schuhmann2023},
positronium \cite{Adkins_2019}, and muonium \cite{crivelli_22}, play a crucial role in testing quantum electrodynamics (QED),
determining fundamental constants, and searching for physics beyond the Standard Model.
All these tasks require accurate theoretical predictions for energy levels of these systems.
If the mass ratio of the two constituent particles is small,
as, e.g., in hydrogen, one can use the Dirac equation as a starting point and
use the QED perturbation theory to account for the recoil and QED corrections.
For systems like positronium and light muonic or antiprotonic atoms, however, the masses of the particles are equal or comparable,
and the Dirac equation is no longer a good approximation.
Thus, one has to rely on the QED formalism from the very beginning in their description.

The QED theory of light atomic systems is based on an expansion in the fine structure constant
$\alpha$ and the derivation of the expansion coefficients as expectation values of various effective
Hamiltonians with the nonrelativistic wave function.
Specifically, the energy of a bound system of two particles
with masses $m_1, m_2$, charges $e_1, e_2$, spins $s_1, s_2$,
and $g$-factors $g_1, g_2$ can be expressed as an expansion
\begin{align}
E(\alpha) =  E^{(0)} + E^{(2)} +  E^{(4)} + E^{(5)} +  E^{(6)} + O(\alpha^7)\,, \label{01}
\end{align}
where the individual terms $E^{(j)} \equiv (Z\,\alpha)^j\,{\cal E}^{(j)}$ are of the order $\alpha^j$.
Here we assume that ${\cal E}^{(i)}$ are real and 
neglect the radiative decay, which induces imaginary corrections to energies.
This effect should be taken into account separately if needed. 
Furthermore, we will exclude from our consideration the vacuum polarization, 
which is either negligible or needs to be taken into account separately, depending on the masses of particles 1 and 2. 
If one of the particles is electron, then the electron vacuum polarization starts at order $\alpha^7$ for $P$-states 
and thus is not relevant for the present study. If both particles are heavier than the electron, 
then the electron vacuum polarization starts at order $\alpha\,(Z\,\alpha)^2$ for $P$-states 
and needs to be accounted for separately, as was done for muonic atoms in Ref.~\cite{rmp}. 
The vacuum polarization with heavier particles in the loop (muons, hadrons) starts at order $\alpha^7$ for $P$-states and is negligible for the present study.
 
Regarding expansion in $\alpha$, the $g$-factor of a particle $a$ defined as
\begin{align}
\vec\mu_a = \frac{e_a\,g_a}{2\,m_a}\,\vec s_a\,, \label{def}
\end{align}
where $\vec{\mu}$ is the magnetic moment, is obtained from experiments. In consequence,  
the $g_a$ factors are not expanded in $\alpha$.  
As a digression we note that this definition in Eq. (\ref{def})
differs from the convention sometimes used in the literature. Specifically, the 
electron $g$-factor is positive, $g=2 + O(\alpha)$, and differs by the sign from 
the definition of Ref.~\cite{codata18}. 
Returning to Eq. (\ref{01}),
the first term of the expansion in $\alpha$ is just
\begin{align}
E^{(0)} = m_1+m_2\,. \label{02}
\end{align}
The next term $E^{(2)}$ is the eigenvalue of the nonrelativistic two-body Hamiltonian $H_0 \equiv H^{(2)}$ in the center-of-mass frame,
\begin{align}
H_0 = \frac{p^2}{2\,\mu} + \frac{e_1\,e_2}{4\,\pi}\,\frac{1}{r}\,, \label{03}
\end{align}
where $\vec p = \vec p_1 = -\vec p_2$, $\vec r = \vec r_1 - \vec r_2$,  and
$\mu = m_1\,m_2/(m_1+m_2)$ is the reduced mass.
If we set $e_1 = -e, e_2 = Z\,e$, the nonrelativistic binding energy becomes
\begin{equation}
E^{(2)} \equiv  E_0 = -\frac{(Z\,\alpha)^2\,\mu}{2\,n^2}\,, \label{04}
\end{equation}
where $n$ is the principal quantum number of the reference state.
The next expansion coefficient,
$E^{(4)}$, is the leading relativistic correction. It is given by the expectation value of the Breit Hamiltonian $H^{(4)}$ \cite{bs}
with the nonrelativistic wave function, $E^{(4)} = \langle H^{(4)} \rangle$,
\begin{widetext}
\begin{align}
H^{(4)} =&\  -\frac{\vec p^{\,4}}{8\,m_1^3}\  -\frac{\vec p^{\,4}}{8\,m_2^3}
+ \frac{e_1\,e_2}{4\,\pi}\,\biggl\{
\frac{1}{2\,m_1\,m_2}\,p^i\,
\biggl(\frac{\delta^{ij}}{r}+\frac{r^i\,r^j}{r^3}\biggr)\,p^j
+\frac{g_1\,g_2}{4\,m_1\,m_2}\,\frac{s_1^i\,s_2^j}{r^3}\,\biggl(\delta^{ij}-3\,\frac{r^i\,r^j}{r^2}\biggr)
\nonumber \\ &
-\frac{\vec r\times\vec p}{2\,r^3} \cdot\biggl[
\frac{g_1}{m_1\,m_2}\,\vec s_1
+\frac{g_2}{m_1\,m_2}\,\vec s_2
+\frac{(g_2-1)}{m_2^2}\,\vec s_2
+\frac{(g_1-1)}{m_1^2}\,\vec s_1\biggr]\biggr\}\,, \label{05}
\end{align}
\end{widetext}
where we assume that the orbital angular-momentum quantum number $l$ of the reference state is positive, $l>0$,
and the spin $s$ of the constituent particles is $0$ or $1/2$.
Let us note that Hamiltonian (\ref{05}) does not account for any annihilation effects, which are present, e.g.,
in positronium.
It also does not include any strong-interaction effects, which are present for hadronic particles.
Such effects, if present, should be evaluated and accounted for separately.
The result for the leading relativistic correction $E^{(4)}$ for a
state with the principal quantum number $n$ and the orbital angular momentum $l=1$ is
\begin{widetext}
\begin{align}
E^{(4)} =&\  \mu^3 (Z\alpha)^4\bigg\{\frac{1}{8\, n^4}\,\biggl(\frac{3}{\mu^2}-\frac{1}{m_1\,m_2}\biggr)
+\frac{1}{6\,n^3}\, \bigg[ -\frac{2}{\mu^2} 
+ \vec L\cdot\vec s_1\bigg(\frac{g_1-1}{m_1^2}+\frac{g_1}{m_1\,m_2}\bigg)
\nonumber\\ &
+ \vec L\cdot\vec s_2\bigg(\frac{g_2-1}{m_2^2} 
+\frac{g_2}{m_1\,m_2}\bigg)
- \frac{3\,g_1\,g_2}{5\,m_1 m_2}s_1^i s_2^j (L^i L^j)^{(2)}\bigg]
\bigg\}\,,\label{06}
\end{align}
\end{widetext}
where the symmetric traceless tensor $(L^i L^j)^{(2)}$ is defined as
\begin{equation}
(L^i L^j)^{(2)} = \frac12\big( L^i L^j + L^j L^i) - \frac{\delta^{ij}}{3}\vec L^2\,. \label{07}
\end{equation}
The QED  correction
of order $\alpha^5$ is denoted by $E^{(5)}$ and given by (for states with $l>0$) \cite{salpeter}
\begin{align}
E^{(5)} =&\
-\frac{14\,(Z\,\alpha)^2}{3\,m_1\,m_2}\,\left\langle
\frac{1}{4\,\pi}\,\frac{1}{r^3}\right\rangle-
\frac{2\,\alpha}{3\,\pi}\,\biggl(\frac{1}{m_1}+\frac{Z}{m_2}\biggr)^2
\nonumber \\ &\ \times
\left\langle\,\vec p\,(H_0-E_0)\,
\ln\bigg[\frac{2\,(H_0-E_0)}{\mu(Z\,\alpha)^2}\bigg]\,\vec{p}\right\rangle\,. \label{08}
\end{align}
The matrix element in the second term is related to the so-called Bethe logarithm $\ln[k_{0}(n, l)]$ by
\begin{align}
\label{09}
	\ln[k_{0}(n, l)] \equiv &\ \frac{n^3}{2 \mu^3 (Z \alpha)^4}
	\nonumber\\& \hspace*{-5ex}\times
	\left\langle \phi\left| \vec{p} \, (H_{0}-E_{0}) \ln\biggl[\frac{2(H_{0}-E_{0})}{\mu (Z\alpha)^2} \biggr] \, \vec{p} \right|\phi \right\rangle\,,
\end{align}
which is tabulated for many hydrogenic states in Ref.~\cite{bethelogs}.
The final result for $E^{(5)}$ for states with $l>0$ is \cite{zatorski_22}
\begin{align}
	E^{(5)} =&\ - \frac{7}{3 \pi} \frac{(Z\alpha)^5 \mu^3}{m_{1} m_{2}} \, \frac{1}{l(l+1)(2l+1)\,n^3} \,
	\nonumber\\ &\
	-\frac{4}{3 \pi} \biggl( \frac{1}{m_1}+\frac{Z}{m_2} \biggr)^2
  \frac{\alpha (Z\alpha)^4 \mu^3}{n^3}\,\ln [k_{0}(n, l)]\,. \label{10}
\end{align}
$E^{(5)}$ is the complete $\alpha^5$ QED correction, provided that the previous-order correction
$E^{(4)}$ is calculated with the physical values of $g$-factors.

\section{NRQED Hamiltonian for the $\bm{\alpha^6}$ correction }
The correction to energy of order $\alpha^6$ can be represented as
\begin{align}
 E^{(6)} = \langle H^{(6)}\rangle + \langle H^{(4)}\,\frac{1}{(E_0-H_0)'}\,H^{(4)}\rangle\,, \label{11}
\end{align}
where the prime in $1/(E_0-H_0)'$ means the exclusion of the reference state from the resolvent, and
$H^{(4)}$ is the Breit-Pauli Hamiltonian given by Eq.~(\ref{05}).
The effective Hamiltonian $H^{(6)}$ can be derived within the framework of nonrelativistic QED (NRQED) \cite{nrqed}.
The starting point of the derivation is
the NRQED Hamiltonian for an arbitrary-spin ($s=0,1/2$) particle, given by ~\cite{ZatorskiPachucki2010}
\begin{widetext}
\begin{align}
	H =&\  eA_{0}
	 + \frac{\vec \pi^{\,2}}{2m}
	 - \frac{e\,g}{2\,m}\, \vec{s}\cdot\vec{B}
	 - \frac{e\,(g-1)}{4\, m^2}\,\vec{s} \cdot (\vec{E}\times\vec{\pi} - \vec{\pi} \times \vec{E} )
	 - \frac{e}{6} \, \Bigl(r_{E}^2+\frac{s\,(s+1)}{m^2}\Bigr) \vec{\nabla}\vec{E}
	  - \frac{e }{120}\,r_{EE}^{4} \nabla^2 \vec{\nabla}\vec{E} \nonumber \\
	&\ - \frac{\vec\pi^{\,4}}{8\, m^3}
	    + \frac{e}{8\,m^3}\Bigl( 2\,\bigl\{\vec \pi^{\,2}, \vec{s}\cdot\vec{B}\bigr\}
           + (g-2)\, \bigl\{\vec{\pi}\cdot\vec{B}, \vec{\pi}\cdot\vec{s} \bigr\} \Bigr)
           - \frac{e}{12\,m} \Bigl(g\,r_{M}^{2} +\frac{3\,(g-2)}{4\,m^2}\Bigr) \vec s\cdot\nabla^2\vec{B} \nonumber \\
	 &\ + \frac{\vec\pi^{\,6}}{16\,m^5}
	  + \frac{e\,(g-1/2)}{24\,m^4}\,s(s+1) \big\{ \vec\pi^{\,2}, \vec{\nabla}\vec{E}\big\}
          + \frac{ie}{32 m^4}\Bigl( 1 +  \frac{s(s+1)}{3} \Bigr) \bigl[\vec\pi^{\,2}, \vec{\pi}\,\vec{E} + \vec{E}\,\vec{\pi}\bigr]	 \nonumber \\
	&\   - \frac{e}{12\,m} \, \Bigl(r_{E}^2 -\frac{g-2}{2\,m^2}\,s\,(s+1)\Bigr)
             \bigl\{\vec{\pi}, \partial_{t}\vec{E} - \vec{\nabla}\times\vec{B}\bigr\}
         + \frac{e\,(g-1/2)}{16 m^4} \,\vec s \, \big\{\vec\pi^{\,2}, \vec{E}\times\vec{\pi} - \vec{\pi} \times \vec{E} \big\} \nonumber\\
    &\   - \frac{e}{24\,m^2} \biggl( g\,r_{M}^{2}  - r_{E}^{2} +\frac{3\,(g-2)}{4\,m^2} \biggr) \vec{s} \,
	  \bigl( \nabla^2 \vec{E}\times\vec{\pi} - \vec{\pi} \times \nabla^2 \vec{E} \bigr)
	  - \frac{e^2}{2} \,\biggl(\alpha_{E}-\frac{s\,(s+1)}{3\,m^2}\biggr)\, \vec E^{\,2}\,, \label{12}
\end{align}
\end{widetext}
where $[X\,,Y] \equiv X\,Y - Y\,X$ denotes the commutator of two operators, and $\{X\,,\,Y\}\equiv X\,Y+Y\,X$ is the anticommutator, $\vec \pi= \vec p-e\,\vec A$.  
In comparison to the original work \cite{ZatorskiPachucki2010} we have redefined the following constants,
\begin{align}
\alpha_E\big|_\mathrm{old} =&\ \alpha_E -\frac{s\,(s+1)}{3\,m^2}\label{13}\,,\\
r_E^2\big|_\mathrm{old} =&\ r_{E}^2+\frac{s\,(s+1)}{m^2} \label{14}\,,\\
r_M^2\big|_\mathrm{old} =&\ \frac{g}{2}\,\Big(r_M^2 + \frac{3}{4\,m^2}\Big)\,, \label{15}
\end{align}
to bring them
in accordance with the standard definitions of the electric dipole polarizability $e^2\,\alpha_E$,
the mean square charge radius $r_E^2 \equiv \lbr r^2\rbr $, and the mean square magnetic radius $r_M^2$.
Furthermore, $r^4_{EE}\equiv \lbr r^4\rbr$ is the mean fourth power of the charge radius.
For the point (scalar or Dirac) particle the parameters are given by
\begin{equation}
r_E^2 = r_{EE}^4 = r_M^2=\alpha_E = g-2=0\,, \label{16}
\end{equation}
whereas for a Dirac particle with the magnetic moment anomaly $\kappa$, they are
\begin{align}
g= &\ 2\,(1+\kappa),\  \,r_E^2=\frac{3\,\kappa}{2\,m^2}, \\
r_M^2 = &\ r_{EE}^4=0,\ \,
\alpha_E =  -\frac{\kappa\,(1+\kappa)}{4\,m^3}\,. \label{17}
\end{align}
For extended-size particles, the parameters $r_E$, $r_M$, and $\alpha_E$
can be in general arbitrary, but we will assume that
$r_E$ and $r_M$ are significantly smaller than the electron Compton wavelength.

\section{Derivation of $\bm{H^{(6)}}$}
Using the NRQED Hamiltonian in Eq.~(\ref{12}), one can derive the effective
operator $H^{(6)}$ for the bound system of two spinless particles, one spinless and one spin-1/2 particle, and two spin-1/2
particles. The derivation follows Ref.~\cite{zatorski_22}, which in turn is based on two former works \cite{Zatorski, nrqed}
and extends the previous calculations of  $H^{(6)}$  to states with $l=1$, where contact terms
contribute. As we will 
show below, the contact terms have previously been accounted for incorrectly
for the positronium $P$-states \cite{Adkins, Czarnecki, Zatorski}.

The typical one-photon exchange contribution between particles $a$ and $b$ is given by
\begin{widetext}
\begin{align}
\langle\phi|\Sigma(E_0)|\phi\rangle =&\ e_a\,e_b\int\frac{d^4
k}{(2\,\pi)^4\,i}\,G_{\mu\nu}(k)\,\biggl\{
\biggl\langle\phi\biggl|\jmath^\mu_a(k)\,e^{i\,\vec k\cdot\vec
r_a} \,\frac{1}{E_0-H_0-k_0+i\,\epsilon}
\,\jmath^\nu_b(-k)\,e^{-i\,\vec k\cdot\vec
r_b}\,\biggr|\phi\biggr\rangle \nonumber \\ &\
+\biggl\langle\phi\biggl|\jmath^\mu_b(k)\,e^{i\,\vec k\cdot\vec
r_b} \,\frac{1}{E_0-H_0-k_0+i\,\epsilon}
\,\jmath^\nu_a(-k)\,e^{-i\,\vec k\cdot\vec
r_a}\,\biggr|\phi\biggr\rangle \biggr\}\,,\label{18}
\end{align}
\end{widetext}
where $G_{\mu\nu}(k)$ is the photon propagator, which is in Feynman gauge $ G^F_{\mu\nu} = g_{\mu\nu}/k^2$,
in Coulomb gauge
\begin{align}
G^C_{\mu\nu}(k) = \left\{
\begin{array}{ll}
{-\frac{1}{\vec{k}^2}} & \mu = \nu = 0\,, \\
{\frac{-1}{k_0^2-\vec k^2 +i\,\epsilon}}\Bigl(\delta_{ij}-{\frac{{k}_i {k}_j}
{\vec{k}^2}\Bigr)} & \mu =i, \, \nu =j\,,
\end{array}
\right. \label{19}
\end{align}
and in temporal gauge
\begin{align}
G^A_{\mu\nu}(k) = \left\{
\begin{array}{ll}
0 & \mu = \nu = 0\,, \\
{\frac{-1}{k_0^2-\vec k^2 +i\,\epsilon}}\Bigl(\delta_{ij}-{\frac{{k}_i {k}_j}
{k_0^2}\Bigr)} & \mu =i, \, \nu =j\,.
\end{array}
\right. \label{20}
\end{align}
The state $\phi$ in Eq.~(\ref{18}) is an eigenstate of $H_0$, and $\jmath^\mu_a$ is the
electromagnetic current operator for particle $a$. The explicit expression for
$\jmath^\mu(k)$ is obtained from the NRQED Hamiltonian in Eq.~(\ref{12})
as the coefficient multiplying
the polarization vector $\epsilon^\mu$ of the electromagnetic potential
\begin{equation}
A^\mu(\vec r,t) \sim \epsilon^\mu_\lambda\, e^{i\,\vec k\cdot\vec r - i\,k_0\,t}\,. \label{21}
\end{equation}
The first terms of the nonrelativistic expansion of the $\jmath^0$ component are
\begin{align}
\jmath^0(k) = &\ 1 +\frac{i\,(g-1)}{2\,m}\,\vec s\cdot\vec
k\times\vec p \nonumber\\ &\ - \frac{1}{6}\,\biggl( r_E^2 +\frac{s\,(s+1)}{m^2}\biggr)\,\vec k^{\,2}+\ldots\label{22}
\end{align}
and those of the $\vec\jmath$ component are
\begin{equation}
\vec \jmath(k) = \frac{\vec p}{m} +
\frac{i\,g}{2\,m}\,\vec s\times\vec k + \ldots\,.\label{23}
\end{equation}
Most of the calculation is performed in the Coulomb gauge in the so-called nonretardation approximation,
in which one sets $k_0=0$ in the photon propagator $G_{\mu\nu}(k)$ and in $\jmath(k)$.
The retardation corrections are considered separately.
Applying the nonretardation approximation and symmetrizing $k_0 \leftrightarrow -k_0$,
the $k_0$ integral in Eq.~(\ref{18}) is evaluated as
\begin{equation}
\frac{1}{2} \int\frac{d\,k_0}{2\,\pi\,i}\,
\biggl[\frac{1}{-\Delta E-k_0+i\,\epsilon}+
\frac{1}{-\Delta E+k_0+i\,\epsilon}\biggr] = -\frac{1}{2}\,, \label{24}
\end{equation}
where we have assumed that $\Delta E$ is positive,  which is the case
when $\phi$ is the ground state. For excited states, the integration
contour is deformed in such a way that all poles from the
electron propagator lie on the same side. Therefore, the result of the $k_0$ integration
for excited states is the same as for the ground state, yielding
\begin{align}
\langle\phi|\Sigma(E_0)|\phi\rangle = &\ -e^2\int\frac{d^3
k}{(2\,\pi)^3}\,G_{\mu\nu}(\vec k)\,
\nonumber\\ &\ \times
\biggl\langle\phi\biggl|\jmath^\mu_a(\vec k)\,e^{i\,\vec
k\cdot(\vec r_a-\vec r_b)} \,\jmath^\nu_b(-\vec
k)\,\biggr|\phi\biggr\rangle\,.\label{25}
\end{align}
The $\vec k$ integral is the Fourier transform
of the photon propagator in the nonretardation approximation
\begin{align}
G_{\mu\nu}(\vec r) = &\ \int \frac{d^3 k}{(2\,\pi)^3} G_{\mu\nu}(\vec
k)\,e^{i\,\vec k\cdot\vec r} \nonumber\\
 = &\ \frac{1}{4\,\pi}\,\left\{
\begin{array}{ll}
-\frac{1}{r}  & \mu = \nu = 0\,, \\
\frac{1}{2\,r}\Bigl(\delta_{ij}+{\frac{{r}_i {r}_j}
{\vec{r}^{\,2}}\Bigr)}& \mu =i, \,\nu =j\,.
 \end{array}
\right.\label{26}
\end{align}
One easily recognizes that  $G_{00}$ is the Coulomb interaction.  
Next-order terms resulting from $\jmath^0$ and $\vec\jmath$ lead to the Breit
Pauli-Hamiltonian, Eq.~(\ref{05}). Below we derive the higher-order
terms in the nonrelativistic expansion, namely the  effective Hamiltonian $H^{(6)}$.
It is expressed as a sum of various contributions
\begin{equation}
H^{(6)} =\sum_{i=0,9} \delta H_i\,.  \label{27}
\end{equation}
We will follow a similar derivation presented in Refs. \cite{zatorski_22, Zatorski, nrqed} for point particles, and use similar notations, namely
 $\vec r = \vec r_1 - \vec r_2$, $e_1 = -e$, $e_2 = Z\,e$, and
the static fields ${\cal A}^0$, $\vec {\cal A}$, and $\vec {\cal E}$ defined as
\begin{align}
e_1\,{\cal A}_1^0 =&\ e_2\,{\cal A}^0_2 = -\frac{Z\,\alpha}{r}\,, \label{28}\\
	e_1 {\cal{A}}^{i}_{1} =& - \frac{Z\alpha}{2\,r} \left( \delta^{ij} + \frac{r^{i} r^{j}}{r^2} \right) \frac{p_{2}^j}{m_2} -
	\frac{Z\,\alpha\,g_2}{2\,m_2} \frac{ \left(\vec{s}_{2} \times \vec{r}\right)^i}{r^3}\,, \label{29} 	\\
e_2 {\cal{A}}^{i}_{2} =& - \frac{Z\alpha}{2\,r} \left( \delta^{ij} + \frac{r^{i} r^{j}}{r^2} \right) \frac{p_{1}^j}{m_1} + \frac{Z\,\alpha\,g_1}{2\,m_1} \frac{ \left(\vec{s}_{1} \times \vec{r}\right)^i}{r^3}\, ,  \label{30}\\
	e_{1}\,\vec{\cal{E}}_1 =& - Z\,\alpha\, \frac{\vec{r}}{r^3},\,\,
	e_{2}\,\vec{\cal{E}}_2 =  Z\,\alpha\, \frac{\vec{r}}{r^3} \,. \label{31}
\end{align}
We now examine the individual contributions $\delta E_i \equiv \langle\delta H_i\rangle $.
$\delta E_0$ is a correction to the kinetic energy,
\begin{align}
	\delta E_{0} =&\ \biggl\langle \frac{p^6}{16\, m_{1}^5} + \frac{p^6}{16\, m_{2}^5}\biggr\rangle \,. \label{32}
\end{align}
$\delta E_1$ is a correction to the Coulomb interaction, where one of the particles interacts by $\delta H$
\begin{widetext}
\begin{align}
	\delta H =&\  - \frac{e }{120}\,r_{EE}^{4}\, \nabla^2 \vec{\nabla}\vec{E}
	  + \frac{e\,(g-1/2)}{24\,m^4}\,s(s+1) \big\{ \vec\pi^{\,2}, \vec{\nabla}\vec{E}\big\}
          + \frac{ie}{32 m^4}\Bigl( 1 +  \frac{s(s+1)}{3} \Bigr) \bigl[\vec\pi^{\,2}, \vec{\pi}\,\vec{E} + \vec{E}\,\vec{\pi}\bigr]	 \nonumber \\
	&\  + \frac{e\,(g-1/2)}{16 m^4} \,\vec s \, \big\{\vec\pi^{\,2}, \vec{E}\times\vec{\pi} - \vec{\pi} \times \vec{E} \big\}
        - \frac{e}{24\,m^2} \biggl( g\,r_{M}^{2}  - r_{E}^{2} +\frac{3\,(g-2)}{4\,m^2} \biggr) \vec{s} \,
	  \bigl( \nabla^2 \vec{E}\times\vec{\pi}
	  - \vec{\pi} \times \nabla^2 \vec{E} \bigr) \,, \label{33}
\end{align}
and the other one by $e\,A^0$. Here we can use the static Coulomb approximation, obtaining
\begin{align}		
	\delta E_{1} =&\ \sum_{a} \biggl\langle- \frac{Z\,\alpha}{8\, m_a^4}\,\biggl(g_a-\frac{1}{2}\biggr)\,  \vec L\cdot\vec{s}_{a}\,\bigg\{ p^{2}, \frac{1}{r^3} \bigg\}
			        + \frac{1}{32 m_a^{4}}\bigg(1+\frac{s_a(s_a+1)}{3}\bigg) \,\Big[p^2, \Big[p^{2}, -\frac{Z\,\alpha}{r} \Big]\Big] \nonumber \\
	                     &+  \frac{Z\alpha }{120}\,r_{EEa}^{4}\,  4\pi\,\nabla^2\delta^3(r)
	                       + i\frac{Z\alpha}{12\,m_a^2} \biggl( g_a\,r_{Ma}^{2}
                               - r_{Ea}^{2} + \frac{3\,(g_a-2)}{4\,m_a^2} \biggr)\,\vec s_a\cdot\vec p\times 4\pi\,\delta^3(r)\,\vec p \biggr\rangle \,, \label{34}
\end{align}
where the second term in Eq. (\ref{33}) vanishes for $l=1$ state.
$\delta E_2$ is a correction to Coulomb interaction when both vertices are
\begin{align}
\delta H = &\ - \frac{e\,(g-1)}{4 m^2}\,\vec{s} \cdot (\vec{E}\times\vec{\pi} - \vec{\pi} \times \vec{E} )
                 - \frac{e}{6} \, \Big( r_{E}^2 + \frac{s\,(s+1)}{m^2}\Big)\, \vec{\nabla}\vec{E}\,. \label{35}
\end{align}
It can also be evaluated in the nonretardation approximation, with the result
\begin{align}	
\delta E_{2} =& \biggl\langle\frac{Z \alpha }{4\,m_{1}^{2} m_{2}^{2}} (g_1-1)(g_2-1)\,
( \vec{s}_2 \times \vec{p})^i \left( \frac{\delta^{ij}}{r^3} - 3 \frac{r^{i} r^{j}}{r^5} +\frac{\delta^{ij}}{3}\,4\,\pi\,\delta^3(r) \right) ( \vec{s}_1 \times \vec{p})^j
\nonumber \\ &
+ \frac{Z\alpha}{36}\,\Big( r_{E1}^2 + \frac{s_1\,(s_1+1)}{m_1^2}\Big)\,\Big( r_{E2}^2 + \frac{s_2\,(s_2+1)}{m_2^2}\Big)\,4\pi\,\nabla^2\,\delta^3(r)
\nonumber \\
&+i\frac{Z\alpha}{12}\biggl[\Big( r_{E1}^2 + \frac{s_1\,(s_1+1)}{m_1^2}\Big)\,\frac{(g_2-1)}{m_2^2}\vec s_2
+ \Big( r_{E2}^2 + \frac{s_2\,(s_2+1)}{m_2^2}\Big)\,\frac{(g_1-1)}{m_1^2}\vec s_1\biggr]\cdot\vec p\times 4\pi\delta^3(r)\,\vec p\biggr\rangle \,. \label{36}
\end{align}
$\delta E_3$ is the relativistic correction to the transverse photon exchange. The first particle is coupled to $\vec A$ by the nonrelativistic term
\begin{align}
\delta H = -\frac{e}{m}\,\vec p\cdot\vec A - \frac{e\,g}{2\,m}\, \vec{s}\cdot\vec{B}\,, \label{37}
\end{align}
and the second one by the relativistic correction
\begin{align}
\delta H = &\
	     \frac{e}{8\,m^3}\Bigl(
           (g-2)\, \bigl\{\vec{\pi}\cdot\vec{B}, \vec{\pi}\cdot\vec{s} \bigr\} %
           +2\,\bigl\{\vec \pi^{\,2}, \vec{s}\cdot\vec{B}\bigr\}\Bigr)
           - \frac{e}{12\, m} \Bigl(g\,r_{M}^{2} +\frac{3\,(g-2)}{4\,m^2}\Bigr) \vec s\cdot\nabla^2\vec{B}
           - \frac{\vec\pi^{\,4}}{8 \,m^3}\,.\label{38}
\end{align}
It is sufficient to calculate it in the nonretardation approximation, which yields
\begin{align}
	\delta E_3 &\ =  \sum_{a}\bigg\langle \frac{1}{4\,m_a^{3}} \bigg(
	\{p^{2} ,\, \vec{s}_a \cdot \vec{\nabla}_a \times e_{a} \vec{\cal{A}}_a\}
	+ \{ p^{2} ,\, \vec{p}_a \cdot e_{a}\vec{\cal{A}}_a\}\bigg)
	+ \frac{e_{a}\,(g_a-2)}{8\,m_a^3} \{\vec{p}_{a} \cdot \vec{\nabla}_{a}\times\vec{\cal{A}}_{a}, \vec{p}_a\cdot \vec{s}_a \}\bigg\rangle	\nonumber \\
& + \bigg\langle
\frac{Z\alpha}{12\, m_1\,m_2} \bigg(g_1\,r_{M1}^{2} +\frac{3\,(g_1-2)}{4\,m_1^2}\bigg)
\Big(i\,\vec s_1\cdot \vec p\times 4\pi\,\delta^3(r)\,\vec p + g_2\,\vec{s}_2 \times \vec{p} \,4\pi\,\delta^3(r)\,\vec{s}_1 \times \vec{p}\Big)
+(1\leftrightarrow2)\bigg\rangle \,. \label{39}
\end{align}
\end{widetext}
$\delta E_4$ comes from the seagull-like coupling
\begin{align}
\delta H = \frac{e^2}{2\,m}\,\vec A^{\,2}\,. \label{40}
\end{align}
Again, the nonretardation approximation yields
\begin{align}
	\delta E_{4} =& \sum_{a} \bigg\langle\frac{e_{a}^2}{2\, m_{a}} \, \vec{\cal{A}}_{a}^{2}\bigg\rangle \, . \label{41}
\end{align}
$\delta E_5$ is a seagull-like term that comes from the coupling
\begin{align}
\delta H =&\   - \frac{e^2}{2} \,\biggl(\alpha_{E}-\frac{s\,(s+1)}{3\,m^2}\biggr)\, \vec E^{\,2} \,, \label{42}
\end{align}
while the second particle is coupled through $e\,A^0$.
It can be obtained in the nonretardation approximation as
\begin{align}
	\delta E_{5} = &\ - \frac{1}{2} \sum_{a} \bigg(\alpha_{Ea}-\frac{s_a(s_a+1)}{3\,m_a^3}\bigg) \, \bigg\langle\frac{Z^2 \alpha^2}{r^4}\bigg\rangle \, . \label{43}
\end{align}
$\delta E_6$ is a seagull-like term that comes from
\begin{align}
\delta H =  - \frac{e\,(g-1)}{4\, m^2}\,\vec{s} \cdot (\vec{E}\times\vec{\pi} - \vec{\pi} \times \vec{E} )\,. \label{44}
\end{align}
Once more the nonretardation approximation can be used, yielding
\begin{align}
	\delta E_6 =& \sum_{a} \, \frac{e_{a}^2\,(g_a-1)}{2\, m_{a}^2} \, \Big\langle\vec{s}_{a} \cdot \vec{\cal{E}}_a \times \vec{\cal{A}}_a\Big\rangle  \, . \label{45}
	\end{align}
$\delta E_7$ is a retardation correction to the single transverse exchange 	
\begin{align}	
		\delta E_{7} =& \,\delta E_{7}^{A} + \delta E_{7}^{B} + \delta E_{7}^{C} \, ,\label{46}
\end{align}
where
\begin{widetext}
\begin{align}
	\delta E_{7}^{A} = &\ \frac{Z\alpha}{16\, m_{1}\,m_{2}} \bigg\langle \frac{2 Z^2 \alpha^2}{r^3} +
	\frac{iZ\alpha r^{i}}{r^3} \bigg[ \frac{p^{2}}{2\,m_{2}}, \frac{r^i r^j - 3 \delta^{ij}\,r^2}{r} \bigg] p^j  \nonumber \\
 &- p^i \bigg[ \frac{r^i r^j - 3 \delta^{ij}\,r^2}{r}, \frac{p^2}{2\,m_{1}} \bigg] \frac{iZ\alpha r^{j}}{r^3}
    - p^{i} \bigg[ \frac{p^2}{2\,m_{2}}, \bigg[ \frac{r^i r^j - 3 \delta^{ij}\,r^2}{r}, \frac{p^2}{2\,m_{1}} \bigg] \bigg] p^j \bigg\rangle + (1 \leftrightarrow 2) \, ,  \label{47} \\	
\delta E_{7}^{B} =&\ \frac{Z \alpha}{8\, m_{1}\,m_{2}} \bigg\langle g_1\,\bigg[ \bigg(\vec{s}_{1}\times \frac{\vec{r}}{r}\bigg)^{i},
	\, \frac{p^2}{2\,m_1} \bigg] \frac{iZ\alpha r^{i}}{r^3}
	- g_2\,\frac{iZ\alpha r^{i}}{r^3} \bigg[ \frac{p^{2}}{2\,m_{2}}, \bigg(\vec{s}_{2}\times \frac{\vec{r}}{r}\bigg)^{i}\bigg]\nonumber\\&
	 + g_1\, \bigg[ \frac{p^2}{2\, m_2}, \bigg[ \bigg(\vec{s}_{1}\times \frac{\vec{r}}{r}\bigg)^{i}, \, \frac{p^2}{2\,m_1} \bigg] \bigg] p^i
	 + g_2\, p^i \bigg[ \frac{p^2}{2\,m_2}, \bigg[ \bigg(\vec{s}_{2}\times \frac{\vec{r}}{r}\bigg)^{i}, \, \frac{p^2}{2\,m_1} \bigg] \bigg] \bigg\rangle
	  + (1 \leftrightarrow 2) \, , \label{48} \\
	\delta E_{7}^{C} =& - \frac{Z\, \alpha\,g_1\,g_2}{16\, m_{1}^{2}\,m_{2}^{2}} \bigg\langle \bigg[ p^2, \bigg[p^2, \, \vec{s}_{1}\vec{s}_{2}\,\frac{2}{3\,r}
	+  s_{1}^{i}\,s_{2}^{j}\frac{1}{2\,r}\bigg( \frac{r^i r^j}{r^2} - \frac{\delta^{ij}}{3} \bigg) \bigg] \bigg] \bigg\rangle \, . \label{49}
	\end{align}
	\end{widetext}
$\delta E_8$ is a retardation correction in a single transverse photon exchange, where one vertex is nonrelativistic, Eq.~(\ref{37}), and the second one is
\begin{align}
\delta H =  - \frac{e\,(g-1)}{4\, m^2}\,\vec{s} \cdot (\vec{E}\times\vec{p} - \vec{p} \times \vec{E} )\,.  \label{50}
\end{align}
The result is
\begin{align}
\delta E_8 =&\ \sum_{a} \biggl\langle \frac{e_{a}^2\,(g_a-1)}{2\, m_{a}^2} \; \vec{s}_a \cdot \vec{\cal{E}}_a \times \vec{\cal{A}}_a
	+ \frac{ie_{a}\,(g_a-1)}{8\,m_{a}^3} \, \nonumber\\ &\
	\times \big[ \, \vec{\cal{A}}_a \cdot(\vec{p}_{a} \times \vec{s}_{a})
	+ (\vec{p}_{a} \times \vec{s}_{a})\cdot \vec{\cal{A}}_a, p_{a}^2 \, \big]\biggr\rangle \, . \label{51}
	\end{align}
The $\delta E_{9}$ contribution arises when one vertex is
\begin{align}
\delta H =&\ - \frac{e}{12\,m} \, \Bigl(r_{E}^2-\frac{g-2}{2\,m^2}\,s\,(s+1)\Bigr)
             \bigl\{\vec{\pi}, \partial_{t}\vec{E} - \vec{\nabla}\times\vec{B}\bigr\}\,,  \label{52}
\end{align}
and the second vertex is nonrelativistic, Eq.~(\ref{37}).
The corresponding current operators are
\begin{align}
\vec j(k) =&\ \frac{\vec p}{m} + \frac{g}{2\,m}\,i\,\vec s\times\vec k\,,  \label{53}\\
\delta j^j(k) =&\ \frac{1}{6\,m}\,\Bigl(r_{E}^2-\frac{g-2}{2\,m^2}\,s\,(s+1)\Bigr)\,p^i\,\big[(\omega^2-\vec k^2)\delta^{ij} \nonumber \\ &\ +k^i\,k^j \big] \,.  \label{54}
\end{align}
For this term we employ the temporal gauge, rather than the Coulomb gauge, and obtain
\begin{align}
\delta E_{9} =& -e_1 e_2 \int\frac{d^3k}{(2\,\pi)^3}\,
\bigg\langle \frac{1}{4}\,\Big\{j_1^i(k)\,,\,\Big\{G_A^{ij}\,\delta j_2^j(-k)\,,\,e^{i\,\vec k\cdot\vec r}\Big\}\Big\}\bigg\rangle \nonumber \\ &\ + (1\leftrightarrow 2)  \label{55}
\end{align}
where
\begin{align}
G_A^{ij}\,\delta j^j(k) =&\ -\frac{1}{6\,m}\,\Bigl(r_{E}^2-\frac{g-2}{2\,m^2}\,s\,(s+1)\Bigr)\,p^i\,.  \label{56}
\end{align}
The result is
\begin{widetext}
\begin{align}
\delta E_{9} = &\
\frac{e_1\,e_2}{6\,m}\,\Bigl(r_{E2}^2-\frac{g_2-2}{2\,m_2^2}\,s_2\,(s_2+1)\Bigr)\,
\biggl\langle
\frac{1}{4}\,\bigg\{\frac{p_1^i}{m_1}\,,\,\bigg\{\frac{p_2^i}{m_2}\,,\,\delta^{3}(r)\bigg\}\bigg\} +
\frac{g_1}{4\,m_1}\,(\vec s_1\times\vec\nabla_1)^i\,\bigg\{
\frac{p_2^i}{m_2}\,,\, \delta^{3}(r)\bigg\}\bigg\rangle + (1\leftrightarrow 2)
\nonumber \\ =&\
\frac{Z\,\alpha}{12\,m_1\,m_2}\, \Bigl(r_{E2}^2-\frac{g_2-2}{2\,m_2^2}\,s_2\,(s_2+1)\Bigr)\,\,\bigg\langle
2\,\pi\, \vec\nabla^2\delta^{3}(r) + i\,g_1\,\vec s_1\cdot\vec p\times 4\,\pi\,\delta^{3}(r)\,\vec p\bigg\rangle + (1\leftrightarrow 2)\,. \label{57}
\end{align}
\end{widetext}
This concludes our derivation of all effective operators to order $\alpha^6$ for $P$-states.
Explicit formulas for matrix elements of elementary and contact operators are presented in Appendix A. 
Matrix elements of other operators can be found in Ref.~\cite{zatorski_22}. 

The last part of $E^{(6)}$ to be evaluated is the second-order iteration of
the Breit Hamiltonian $H^{(4)}$ in Eq.~(\ref{11}).
It  has already been derived for arbitrary $l>0$ in Ref.~\cite{zatorski_22} by the method developed in Ref.~\cite{Zatorski}, 
and the result is valid also for the case $l=1$ investigated here, provided the contribution from intermediate states with $l-2$ 
is explicitly excluded. Since the derivation and the final expressions
are quite long, we refer the reader to Ref.~\cite{zatorski_22} for the corresponding formulas.

Adding together all contributions, we arrive at our final result for the
$\alpha^6$ correction for $nP$ states. It is written as $E^{(6)} = (Z\,\alpha)^6\,{\cal E}^{(6)}$,
\begin{align}
{\cal E}^{(6)} =&\  {\cal E}_{NS} + \vec s_1\cdot\vec s_2\,{\cal E}_{SS}
                            + \vec L\cdot\vec s_1\,{\cal E}_{L1}
                            + \vec L\cdot\vec s_2\,{\cal E}_{L2} \nonumber \\ &\
                            + (L^i\,L^j)^{(2)}\,s_1^i\,s_2^j\,{\cal E}_{LL} \,, \label{58}
\end{align}
where
\begin{align}
{\cal E}_{NS} =&\ {\cal E}_{S0} + \frac{4}{3}\,s_1\,(s_1+1)\,{\cal E}_{S1} + \frac{4}{3}\,s_2\,(s_2+1)\,{\cal E}_{S2} \nonumber \\ &\
 + \frac{16}{9}\,s_1\,(s_1+1)\,s_2\,(s_2+2)\,{\cal E}_{S12} \,, \\
{\cal E}_{L1} =&\  {\cal E}_{LN1}  +\frac{4}{3}\,s_2(s_2+1)\,{\cal E}_{LS1}\,, \\
{\cal E}_{L2} =&\ {\cal E}_{LN2} +\frac{4}{3}\,s_1\,(s_1+1)\,{\cal E}_{LS2} \,,
\end{align}
with the individual terms given by
\begin{widetext}
\begin{align}
{\cal E}_{S0} =&\ \mu\biggl(-\frac{5}{16\,n^6} + \frac{1}{2\,n^5} - \frac{1}{6\,n^4} - \frac{1}{27\,n^3}\biggr) +
 \frac{\mu^3}{m_1\,m_2}\biggl(\frac{3}{16\,n^6} - \frac{13}{30\,n^5} + \frac{2}{5\,n^3}\biggr)
 - \frac{\mu^5}{m_1^2\,m_2^2}\,\frac{1}{16\,n^6}
 \nonumber \\ &\
 +\mu^5\,\biggl(\frac{1}{n^3}-\frac{1}{n^5}\biggr)\biggl( \frac{2}{27}\,r^2_{E1}\,r^2_{E2}
 + \frac{r^2_{E1} + r^2_{E2}}{9\,m_1\,m_2}
  + \frac{r^4_{EE1} + r^4_{EE2}}{45}\biggr)
  - \mu^4\,\frac{\alpha_{E1} + \alpha_{E2}}{5}\,\biggl(\frac{1}{n^3}-\frac{2}{3\,n^5}\biggr)\,,
\\[2ex]
{\cal E}_{S2} =&\
\frac{\mu^3}{m_2^2}\,\frac{g_2^2}{24}\,\biggl(\frac{1}{5\,n^5} - \frac{1}{2\,n^4} - \frac{119}{180\,n^3}\biggr)
+ \frac{\mu^5}{m_2^4}\,\biggl[\frac{g_2}{24}\,\biggl(\frac{1}{n^3}-\frac{1}{n^5}\biggr) + \frac{7}{60\,n^5} - \frac{1}{48\,n^4} - \frac{641}{4320\,n^3}\biggr]
\nonumber \\ &\
+  \frac{\mu^4}{m_2^3}\,\biggl[ -\frac{g_2^2}{40}\,\biggl( \frac{1}{n^3} - \frac{2}{3\,n^5}\biggr)
+ \frac{g_2}{24}\,\biggl(-\frac{1}{5\,n^5} + \frac{1}{n^4} + \frac{137}{90\,n^3}\biggr) - \frac{7}{60\,n^5} +  \frac{2}{15\,n^3}\biggr]
+ \frac{\mu^5}{m_2^2}\,\frac{r_{E1}^2}{18}\,\biggl(\frac{1}{n^3}-\frac{1}{n^5}\biggr)\,,
\\[2ex]
	{\cal E}_{S12}=&\ \frac{\mu^5}{m_1^2\,m_2^2}\,\biggl[\biggl(\frac{5}{3\,n^5}- \frac{1}{n^4} - \frac{287}{90\,n^3}\biggr)\,\frac{g_1^2\,g_2^2}{640}
+\frac{1}{24}\,\biggl(\frac{1}{n^3}-\frac{1}{n^5}\biggr) \biggr]\,,\\[2ex]
{\cal E}_{LN2} =&\
\frac{\mu^2}{m_2}\,g_2\,\biggl(-\frac{1}{3\,n^5} + \frac{1}{6\,n^4} + \frac{13}{108\,n^3}\biggr)
 +  \frac{\mu^3}{m_2^2}\,\biggl[
  g_2^2\,\biggl(-\frac{1}{40\,n^5} + \frac{1}{48\,n^4} + \frac{227}{4320\,n^3}\biggr)
 + g_2\,\biggl(\frac{3}{10\,n^5} - \frac{1}{5\,n^3}\biggr)
 \nonumber \\ &\
 + \frac{5}{12\,n^5} - \frac{1}{6\,n^4} - \frac{13}{108\,n^3}\biggr]
+ \frac{\mu^4}{m_2^3}\,\biggl[g_2\,\biggl(-\frac{1}{6\,n^5} - \frac{1}{24\,n^4} + \frac{5}{432\,n^3}\biggr) - \frac{5}{12\,n^5} + \frac{1}{6\,n^3}\biggr]
    \nonumber \\ &\
+ \frac{\mu^5}{m_2^4}\,\biggl[ \frac{1}{4\,n^5} + \frac{1}{48\,n^4} - \frac{41}{864\,n^3}\biggr]
 + \frac{1}{9}\,\biggl(\frac{1}{n^3}-\frac{1}{n^5}\biggr)\,\biggl[
 \biggl( -\frac{\mu^4}{m_2}\,g_2 + \frac{\mu^5}{m_2^2} \biggr)\,r^2_{E1}
 + \frac{\mu^5}{m_2^2}\,r^2_{E2}
 - \frac{\mu^4}{m_2} \,g_2\,r^2_{M2}\biggr]\,,
 \\
{\cal  E}_{LS2}=&\
\frac{\mu^4}{m_1^2\,m_2}\,\frac{g_2}{12}\,\biggl[ \frac{1}{n^5}-\frac{1}{n^3}
- g_1\,\biggl(\frac{7}{20\,n^5} + \frac{1}{8\,n^4} - \frac{133}{720\,n^3}\biggr)
+ g_1^2\,\biggl(-\frac{3}{20\,n^5} + \frac{1}{8\,n^4} + \frac{227}{720\,n^3}\biggr) \biggr]
\nonumber \\ &\
+ \frac{\mu^5}{m_1^2\,m_2^2}\,\frac{1}{12}\,\biggl[ \frac{1}{n^3}-\frac{1}{n^5}
+g_1\,g_2\,\biggl( \frac{7}{20\,n^5} + \frac{1}{8\,n^4} - \frac{133}{720\,n^3}\biggr) +
g_1^2\,g_2^2\,\biggl(-\frac{1}{40\,n^5} + \frac{9}{320\,n^4} + \frac{187}{3200\,n^3}\biggr)\biggr]\,,
  \\[2ex]
   {\cal E}_{SS} =&\
  -\frac{\mu^3}{m_1\,m_2}\,g_1\,g_2\,\biggl( \frac{1}{60\,n^5} + \frac{1}{18\,n^4}  + \frac{47}{1620\,n^3}\biggr) 
  + \frac{\mu^4}{m_1\,m_2}\,\biggl(\frac{g_1}{m_2} + \frac{g_2}{m_1}\biggr)\,\biggl( \frac{1}{18\,n^5} + \frac{1}{18\,n^4} - \frac{5}{324\,n^3}\biggr)
  \nonumber \\ &\
  +\frac{\mu^5}{m_1^2\,m_2^2}\,\biggl[
  \frac{g_1^2\,g_2^2}{480}\,\biggl(\frac{5}{3\,n^5} - \frac{1}{n^4} - \frac{287}{90\,n^3}\biggr)
  +\frac{1}{30\,n^5}  - \frac{1}{18\,n^4}  - \frac{191}{1620\,n^3} \biggr] \nonumber \\
  &\
  +  \frac{2}{27}\biggl(\frac{1}{n^3}-\frac{1}{n^5}\biggr)\frac{\mu^5}{m_1\,m_2}\,g_1\,g_2\,(r^2_{M1} + r^2_{M2})\,,
\\
	{\cal E}_{LL} =&\
\frac{\mu^3}{m_1\,m_2}\,\frac{g_1\,g_2}{4}\,\biggl( \frac{51}{50\,n^5} - \frac{7}{12\,n^4} - \frac{3697}{5400\,n^3}\biggr)
+ \frac{\mu^4}{m_1\,m_2} \biggl[ \biggl(\frac{g_1}{m_1}+\frac{g_2}{m_2}\biggr)\,
g_1\,g_2\,\biggl( \frac{9}{200\,n^5} - \frac{3}{80\,n^4} - \frac{227}{2400\,n^3}\biggr)
\nonumber \\ &\
+  \biggl(\frac{g_1}{m_2}+\frac{g_2}{m_1}\biggr)\,\biggl(-\frac{19}{150\,n^5} + \frac{1}{12\,n^4} + \frac{1171}{5400\,n^3}\biggr)\biggr]
+ \frac{\mu^5}{m_1^2\,m_2^2}\,\biggl[
\frac{g_1^2\,g_2^2}{200}\,\biggl(\frac{7}{6\,n^5} - \frac{7}{8\,n^4} - \frac{1709}{720\,n^3}\biggr)
\nonumber \\ &\
+ g_1\,g_2\,\biggl(-\frac{6}{25\,n^5} - \frac{3}{40\,n^4} + \frac{37}{1200\,n^3}\biggr)
-\frac{g_1 + g_2}{10}\,\biggl( \frac{1}{n^3} - \frac{1}{n^5}\biggr)
+ \frac{2}{25\,n^5} - \frac{1}{12\,n^4} - \frac{1063}{5400\,n^3} \biggr]
\nonumber \\ &\
+ \frac{\mu^5}{m_1\,m_2}\,\frac{g_1\,g_2}{9}\,\biggl(  \frac{1}{n^3}  -\frac{1}{n^5}\biggr)\,(r^2_{M1} +  r^2_{M2})\,.
\end{align}
\end{widetext}
We remind the reader that $E^{(6)}$ is the complete $\alpha^6$ QED correction, provided that the lower-order correction
$E^{(4)}$ is calculated with the physical values of $g$-factors.

We now turn to the comparison of the obtained formulas for the $l = 1$ states
with the general $l>0$ result of Ref.~\cite{zatorski_22}
derived with the omission of contact terms. The contact terms vanish in the $l > 1$ case
but are present for $l = 1$ (even for the point particles).
We will consider separately the cases of two spinless particles, of one spinless and one spin-$1/2$ particle,
and of two spin-$1/2$ particles.

\section{Spin $\bm {s_1=s_2=0}$}
For a system consisting of two spinless particles, $s_1=s_2=0$,  ${\cal E}^{(6)} =  {\cal E}_{S0}$.
This result differs from the general result ${\cal E}^{(6)}_G$ from Ref. \cite{zatorski_22} by the finite-size terms only,
as it should,
\begin{align}
 {\cal E}^{(6)}  - {\cal E}^{(6)}_G\bigr|_{l=1} =&\
\frac{\mu^5}{9}\,\biggl(\frac{1}{n^3}-\frac{1}{n^5}\biggr)\biggl( \frac{2}{3}\, r^2_{E1}\, r^2_{E2} \nonumber \\ &\
 +\frac{ r^2_{E1} +  r^2_{E2}}{m_1\,m_2}  + \frac{r^4_{EE1} + r^4_{EE2}}{5}\biggr)\,. \label{60}
\end{align}
In the infinite-mass limit of one of the particles (and only in this limit), ${\cal E}^{(6)}$ corresponds to a solution of the Klein-Gordon equation.
For an arbitrary mass ratio there is no fundamental equation and energy levels can be obtained only from the QED theory.

An example of a bound system of two scalar particles is the pionic helium atom
investigated by Masaki Hori \cite{hori_20, hori_21}.
However, in this case
the short-range interactions are dominated by strong forces, and the above formula thus has limited applicability.

\section{Spin $\bm{s_1=0, s_2=1/2}$}
For a system consisting of particles with $s_1=0$ and $s_2=1/2$, the binding energy at the order $\alpha^6$ is
\begin{align}
{\cal E}^{(6)} =  {\cal E}_{S0} + {\cal E}_{S2} +  \vec L\cdot\vec s_2\,{\cal E}_{LN2}\,. \label{61}
\end{align}
The difference of ${\cal E}^{(6)}$ and the general result ${\cal E}_G$ from Ref. \cite{zatorski_22} is
\begin{align}
{\cal E}^{(6)} -{\cal E}^{(6)}_G\bigr|_{l=1} =&\
\biggl[ \frac{r^2_{E1}+\tilde r^2_{E2}}{m_1\,m_2}+ \frac{2}{3}\,\,r^2_{E1}\,\biggl(r^2_{E2}+\frac{3}{4\,m_2^2}\biggr)
\nonumber \\ &\ \hspace{-15ex}
+ \frac{r^4_{EE1}+ r^4_{EE2}}{5}
+  \vec L\cdot\vec s_2\,\biggl(
 -\frac{g_2\,(\tilde r^2_{M2}+r^2_{E1})}{m_1\,m_2}
 \nonumber \\ &\ \hspace{-15ex}
 + \frac{r^2_{E2} - g_2\,\tilde r^2_{M2} - (g_2-1)\,r^2_{E1}}{m_2^2}
\biggr)\biggr]\,\frac{\mu^5}{9}\,\biggl(\frac{1}{n^3}-\frac{1}{n^5}\biggr)\,, \label{62}
\end{align}
where
\begin{align}
g\,\tilde r^2_M =&\ g\,r^2_{M} + \frac{3\,(g-2)}{4\,m^2}\,,\\
\tilde r^2_E =&\ r^2_E -\frac{3\,(g-2)}{8\,m^2}\,.
\end{align}
With the rotational angular momentum $l=1$ coupled to the spin $s_2=1/2$, the total angular momentum $J$ can be either $J =1/2$ or $3/2$. The corresponding energies are
\begin{align}
{\cal E}^{(6)}\big|_{J=1/2} =&\  {\cal E}_{S0} + {\cal E}_{S2} -  {\cal E}_{LN2}\,, \label{63} \\
{\cal E}^{(6)}\big|_{J=3/2} =&\  {\cal E}_{S0} + {\cal E}_{S2} + \frac12 {\cal E}_{LN2}\,. \label{64}
\end{align}

The explicit formulas for ${\cal E}^{(6)}$ are quite long. However, their expansion for a small
mass ratio $m_2/m_1$ is quite compact.
Specifically,
assuming that particle 2 is point-like $(\kappa_2 = r^2_{E2} = r^2_{M2} = r^4_{EE2} = 0)$
and neglecting the polarizabilities $(\alpha_{E1}=\alpha_{E2} = 0)$,  we obtain,
${\cal E}^{(6)} = {\cal E}^{(6,0)} + {\cal E}^{(6,1)} + \ldots\,$,
\begin{widetext}
\begin{align}
{\cal E}^{(6,0)}\big|_{J=1/2} =&\  m_2\,\biggl[\biggl(-\frac{5}{16\,n^6} + \frac{3}{4\,n^5} - \frac{3}{8\,n^4} - \frac{1}{8\,n^3}\biggr) +
 \frac{1}{6}\,\biggl(\frac{1}{n^3}-\frac{1}{n^5}\biggr)\,m_2^2\,r^2_{E1} + \frac{1}{45}\,\biggl(\frac{1}{n^3}-\frac{1}{n^5}\biggr)\,m_2^4\,r^4_{EE1}\biggr]\,, \label{65}
  \\
{\cal E}^{(6,1)}\big|_{J=1/2} =&\ \frac{m_2^2}{m_1}\,\biggl[
 \biggl(\frac{1}{2\,n^6} - \frac{19}{15\,n^5} + \frac{3}{8\,n^4} + \frac{21}{40\,n^3}\biggr)
 -\frac{1}{2}\,\biggl(\frac{1}{n^3}-\frac{1}{n^5}\biggr)\,m_2^2\,r^2_{E1} -\frac{1}{9}\,\biggl(\frac{1}{n^3}-\frac{1}{n^5}\biggr)\,m_2^4\,r^4_{EE1} \biggr] \,, \label{66}
 \\
{\cal E}^{(6,0)}\big|_{J=3/2} =&\
m_2\,\biggl[\biggl(-\frac{5}{16\,n^6} + \frac{3}{8\,n^5} - \frac{3}{32\,n^4} - \frac{1}{64\,n^3}\biggr) + \frac{1}{45}\,\biggl(\frac{1}{n^3}-\frac{1}{n^5}\biggr)\,m_2^4\,r^4_{EE1}\biggr]\,, \label{67}
\\
{\cal E}^{(6,1)}\big|_{J=3/2} =&\ \frac{m_2^2}{m_1}\,\biggl[
\biggl(\frac{1}{2\,n^6} - \frac{23}{30\,n^5} + \frac{3}{32\,n^4} + \frac{133}{320\,n^3}\biggr)
 -\frac{1}{9}\,\biggl(\frac{1}{n^3}-\frac{1}{n^5}\biggr)\,m_2^4\,r^4_{EE1}\biggr]\,. \label{68}
\end{align}
\end{widetext}
In the point-nucleus limit, these formulas
are in agreement with the literature results \cite{codata18}.
Furthermore, the finite-size corrections in the nonrecoil limit agree with those derived in Ref.~\cite{three}.
The finite-size recoil corrections are a new result obtained here.
We have verified it by comparing with numerical calculations performed to all orders in $\Za$ in Sec.~\ref{sec viii}.

\section{Spin $\bm{s_1=s_2=1/2}$}

The most complicated case considered here
is when both particles have spin $s=1/2$. The binding energy ${\cal E}^{(6)}$ can then be expressed as
\begin{widetext}
\begin{align}
{\cal E}^{(6)} =&\  {\cal E}_{S0} + {\cal E}_{S1} + {\cal E}_{S2} + {\cal E}_{S12} + \vec s_1\cdot\vec s_2\,{\cal E}_{SS}
+ \vec L\cdot\vec s_1\,({\cal E}_{LN1} + {\cal E}_{LS1})  + \vec L\cdot\vec s_2\,({\cal E}_{LN2}+{\cal E}_{LS2})
+ (L^i\,L^j)^{(2)}\,s_1^i\,s_2^j\,{\cal E}_{LL} \,.
\label{69}
\end{align}
It differs from the general result ${\cal E}_G$ from Ref.~\cite{zatorski_22} by
\begin{align}
		{\cal E}^{(6)}-{\cal E}^{(6)}_G\big|_{l=1} =&\ \frac{\mu^5}{9}\,\biggl( \frac{1}{n^3} - \frac{1}{n^5}\biggr) \biggl\{
\frac{\tilde r^2_{E1} + \tilde r^2_{E2}}{m_1\,m_2}
+ \frac{2}{3}\,r^2_{E1}\,r^2_{E2} + \frac{r^2_{E1}}{2\,m_2^2} + \frac{r^2_{E2}}{2\,m_1^2}
+ \frac{r^4_{EE1}+r^4_{EE2} }{5} + \frac{3}{8\,m_1^2\,m_2^2}\bigg(1-\frac{g_1^2\,g_2^2}{16}\bigg)
\nonumber \\ &\
+ \vec L\cdot\vec s_1\,\biggl[ -\frac{g_1\,\bigl(\tilde r^2_{M1}  + \tilde r^2_{E2} \bigr)}{m_1\,m_2}
+\frac{r^2_{E1} + (1-g_1) r^2_{E2}  - g_1\,\tilde r^2_{M1}}{m_1^2} +\frac{3}{4\,m_1^2\,m_2^2}\bigg(2-g_1 + \frac{g_1^2\,g_2^2-16}{16}\bigg)\biggr]
\nonumber \\ &\
+ \vec L\cdot\vec s_2\, \biggl[
-\frac{g_2\,\bigl(\tilde r^2_{M2}  + \tilde r^2_{E1}\bigr)}{m_1\,m_2}
+\frac{(1-g_2)r^2_{E1} + r^2_{E2} - g_2\,\tilde r^2_{M2}}{m_2^2} 
+\frac{3}{4\,m_1^2\,m_2^2}\bigg(2-g_2 + \frac{g_1^2\,g_2^2-16}{16}\bigg) \biggr]
\nonumber \\ &\
+ \vec s_1\cdot\vec s_2\,\biggl[
\frac{2}{3}\,\frac{g_1\,g_2}{m_1\,m_2} \,(\tilde r^2_{M1} + \tilde r^2_{M2})
- \frac{1}{4\,m_1\, m_2}\bigg(g_2\,\frac{g_1-2}{m_1^2} + g_1\,\frac{g_2-2}{m_2^2} \bigg)\nonumber \\
&\ - \frac{(g_1-2)^2\,(g_2-2)^2 + 4\,(g_1-2)^2 + 4\,(g_2-2)^2 + 4\,(g_1-2)(g_2-2)(g_1+g_2-4)}{32\,m_1^2\,m_2^2}  \biggr]
\nonumber \\ &\
+ (L^i\,L^j)^{(2)}\,s_1^i\,s_2^j\,\biggl[
\frac{g_1\,g_2}{m_1\,m_2}\,\bigl( \tilde r^2_{M1}  +  \tilde r^2_{M2}\bigr)
+ \frac{3}{10\,m_1\, m_2}\bigg(g_2\,\frac{g_1-2}{m_1^2} + g_1\,\frac{g_2-2}{m_2^2} \bigg) \nonumber \\
&\
- \frac{3\,\big[(g_1-2)^2\,(g_2-2)^2 + 4\,(g_1-2)^2 + 4\,(g_2-2)^2 + 4\,(g_1-2)(g_2-2)(g_1+g_2-4)\big]}{16\,m_1^2\,m_2^2} \biggr] \biggr\}\,. \label{70}
\end{align}
\end{widetext}
The above difference vanishes in the point-particle limit, which indicates an agreement
not only with Ref.~\cite{zatorski_22} but also with previous calculations \cite{Zatorski, Adkins, Czarnecki}
since Ref.~\cite{zatorski_22} was claimed to agree with them in the limit $m_1=m_2$.
However, this agreement with Ref.~\cite{zatorski_22} is due to an accidental cancellation of two mistakes.
The first one is the neglect of local term in calculating $\delta E_2$ for $l =1$ state in Ref.~\cite{zatorski_22}.
On closer inspection,  Zatorski calculates it for point particles \cite[Eq. (94)]{Zatorski},
separately for the $l=1$ case \cite[Eq. (99)]{Zatorski}  and for the $l>1$ case \cite[Eq.~(103)]{Zatorski}, 
closely following the original calculation of Khriplovich \cite{Khriplovich, Golosov}.
Later he writes that ``\ldots the correction $\delta E_2$ for $l =1$ still can be obtained from Eq. (103)", which we find to be incorrect.
The second mistake is the implicit inclusion of intermediate states with $l-2$ angular momentum, which should vanish for $l=1$.
Surprisingly, these two mistakes cancel each other, leading to correct result for point particles with $g=2$.
We conclude that although the case $l=1$ requires separate treatment
compared to the $l>1$ states, the final formula presented in Ref.  \cite{zatorski_22}, originally derived for $l>1$, remains valid also for $l=1$.
Therefore, the previous result for the positronium $P$-levels reported in the literature \cite{Czarnecki, Zatorski, Adkins} 
is correct, and for readers convenience the corresponding  formula is presented in Appendix~\ref{sec:app2}.

Returning to Eq.~(\ref{69}), we present formulas for
its expansion in the small mass ratio $m_2/m_1$,
for the case of the point-like second particle $(\kappa_2 = r^2_{E2} = r^2_{M2} = r^4_{EE2}  = 0)$
and negligible polarizabilities $(\alpha_{E1}=\alpha_{E2} = 0)$. The results are
\begin{widetext}
\begin{align}
{\cal E}^{(6,0)} =&\ m_2\,\biggl\{
\biggl(- \frac{5}{16\,n^6} + \frac{1}{2\,n^5} - \frac{3}{16\,n^4} - \frac{5}{96\,n^3}\biggr)
+\vec L\cdot\vec s_2\,\biggl(-\frac{1}{4\,n^5} + \frac{3}{16\,n^4} + \frac{7}{96\,n^3}\biggr)
\nonumber \\ &\
+ \frac{1}{9}\,\biggl(\frac{1}{n^3}-\frac{1}{n^5}\biggr)\,\biggl[ \biggl(\frac{1}{2} - \vec L\cdot\vec s_2\biggr)\,m_2^2\,r^2_{E1}
+ \frac{1}{5}\,m_2^4\,r^4_{EE1}\biggr]\biggr\}\,,
\label{71} \\
{\cal E}^{(6,1)} =&\
\frac{m_2^2}{m_1}\,\biggl\{ \biggl(\frac{1}{2\,n^6} - \frac{14}{15\,n^5} + \frac{3}{16\,n^4} + \frac{217}{480\,n^3}\biggr)
+ \vec L\cdot \vec s_2\,\biggl(\frac{1}{3\,n^5} - \frac{3}{16\,n^4} - \frac{7}{96\,n^3}\biggr)
\nonumber \\ &\
+ g_1\,\vec L\cdot \vec s_1\,\biggl(-\frac{43}{120\,n^5} + \frac{3}{16\,n^4} + \frac{83}{480\,n^3}\biggr)
+ g_1\,\vec s_1\cdot\vec s_2\,\biggl(\frac{1}{45\,n^5} - \frac{1}{18\,n^4} - \frac{119}{1620\,n^3}\biggr)
\nonumber \\ &\
+ g_1\,(L^i\,L^j)^{(2)}\,s_1^i\,s_2^j\,\biggl(\frac{169}{300\,n^5} - \frac{43}{120\,n^4}
 - \frac{5441}{10800\,n^3}\biggr)
+\frac{1}{9}\,\biggl(\frac{1}{n^3}-\frac{1}{n^5}\biggr)\,\biggl[ -m_2^4\,r^4_{EE1} \nonumber \\ &\
-  3\,\bigg(\frac{1}{2}-\vec L\cdot\vec s_2\bigg)\,m_2^2\,r^2_{E1}
+ \biggl( \frac{4}{3}\,\vec s_1\cdot\vec s_2 -\vec L\cdot\vec s_1
+ 2\,(L^i\,L^j)^{(2)}\,s_1^i\,s_2^j\biggr)\, m_2^2\,g_1\,r^2_{M1} \biggr]\biggr\}\,. \label{72}
\end{align}
\end{widetext}
The above expression for ${\cal E}^{(6,0)}$ agrees with that for the $s_1=0, s_2=1/2$ case, as it should.
Similarly, ${\cal E}^{(6,1)}$ agrees with the $s_1=0, s_2=1/2$ case up to the terms with $\vec s_1$.
The $\vec s_1$-dependent terms are responsible for the hyperfine structure at the $\alpha^6$ order
and for mixing of the $P_{1/2,F=1}$ and $P_{3/2,F=1}$ states.

\section{$\bm{2P}$ fine structure in light muonic atoms}

Accurate theoretical predictions of the fine and hyperfine structure of the $2P$ levels in muonic atoms
are required
for the determination of the nuclear charge radii from experimental $2P$-$2S$ transition energies.
QED calculations of the $2P$ fine structure of $\mu$He ions have been performed in Refs.~\cite{sgk2017, korzinin2018},
neglecting higher-order terms in the mass ratio, namely  $(Z\,\alpha)^6\,m_\mu\,(m_\mu/m_N)^{(2+)}$,
where the subscripts ${\mu}$ and $N$ refer to the muon and the nucleus, respectively. In the present work
we obtain the result for the $\alpha^6$ contribution with full dependence on the mass ratio $m_\mu/m_N$.

\begin{table}
\caption{$2P$ fine structure  of $\mu$He ions, in meV.
The root-mean-square nuclear-charge radii are \cite{schuhmann2023, rmp} $r_E(h) = 1.970$~fm, $r_E(\alpha) = 1.679$~fm.
Our uncertainty is due to higher-order in $\alpha$ terms, mainly due to the  two-loop electron vacuum polarization.
From the previous results in Ref. \cite{sgk2017, pohl2016}  we have subtracted  BP(tot) $=  0.1947$ meV  due to a different definition of the fine structure of $\mu^3$He
used in these works.}
\label{fsmu}
\begin{center}
\begin{tabular}{ld{4.12}d{4.12}}
    \hline \hline & \\
\centt{contribution} & \centt{$\mu^3$He$^+$} &  \centt{$\mu^4$He$^+$}  \\[1ex]
\hline
$E^{(4)}_\mathrm{fs}$ & 144.510\,95 &  145.898\,24   \\
$E^{(4)}_\mathrm{fs,vp} $ & 0.269\,81& 0.275\,65    \\
$E^{(6)}_\mathrm{fs}$ &  0.004\,19 & 0.007\,64  \\[2ex]
$E_\mathrm{fs}$ & 144.785(3) &   146.182(3)  \\
Refs. \cite{sgk2017, korzinin2018}  & 144.785(5)  &  146.181(5) \\
exp. \cite{pohl2016, schuhmann2023}  &144.763(114) & 146.047(96)  \\

\hline\hline&
\end{tabular}
\end{center}
\end{table}

The binding energy of a muonic atom can be decomposed in terms of basic angular-momentum operators,
similarly to Eq.~(\ref{58}),
\begin{align}
E =&\  E_{NS} + \vec L\cdot\vec s_\mu\,E_{L\mu} +  \vec L\cdot\vec s_N\,E_{LN} \nonumber \\ &\
+ (L^i\,L^j)^{(2)}\,s_N^i\,s_\mu^j\,E_{LL} + \vec s_N\cdot\vec s_\mu\,E_{SS}\,. \label{gen1}
\end{align}
Here, the spin-independent term $E_{NS}$ corresponds to the energy centroid,
the second term is responsible for the fine splitting,
$E_\mathrm{fs} \equiv 3/2\,E_{L\mu}$, whereas the remaining terms
induce the hyperfine splitting and mixing between the fine and hyperfine structure.

We are now interested in the fine structure of the $2P$ state.
The leading fine structure of order $(Z\,\alpha)^4$ is obtained from Eq.~(\ref{06}),
with the result
\begin{align}
E^{(4)}_\mathrm{fs} =&\  \frac{\mu^3 (Z\alpha)^4}{32}\,\bigg(\frac{g_\mu-1}{m_\mu^2}+\frac{g_\mu}{m_\mathrm{N}\,m_\mu}\bigg)\,. \label{78}
\end{align}
For the $\alpha^6$ correction, we set $g_\mu=2$ because
the magnetic-moment anomaly is only a part of the $\alpha^7$ correction.
Similarly, we neglect QED corrections to $r^2_E$ and $r^2_M$ of the muon.
We obtain for the $s_N=0$ nucleus
\begin{widetext}
 \begin{align}
E^{(6)}_\mathrm{fs} =&\  \frac{3}{2}\,E_{L\mu}(n=2,g_\mu=2, s_N=0)\nonumber \\
=&\ \mu\,\frac{(Z\,\alpha)^6}{64}\,\biggl[
\frac{5}{4} + \frac{1}{4}\,\frac{\mu}{m_N}
- \frac{19}{18}\,\Big(\frac{\mu}{m_N}\Big)^2
- \frac{3}{4}\, \Big(\frac{\mu}{m_N}\Big)^3
+  \frac{11}{36}\,\Big(\frac{\mu}{m_N}\Big)^4
 - \mu^2\,r^2_{E}\,\Big(1-\frac{\mu^2}{m_N^2}\Big)\biggr]\,, \label{79}
\end{align}
whereas for $s_N=1/2$
\begin{align}
	E^{(6)}_\mathrm{fs} =&\ \frac{3}{2}\,E_{L\mu}(n=2, g_\mu=2, s_N = 1/2) \nonumber \\
=&\ \mu\,\frac{(Z\,\alpha)^6}{64}\, \biggl[
\frac{5}{4} + \frac{1}{4}\,\frac{\mu}{m_N}
+\bigg(-\frac{19}{18} + \frac{851}{900}\,g_N^2\bigg)\,\Big(\frac{\mu}{m_N}\Big)^2
+  \bigg(-\frac{3}{4} + \frac{5}{72}\,g_N - \frac{2179}{1800}\,g_N^2\bigg)\,\Big(\frac{\mu}{m_N}\Big)^3
\nonumber \\ &\
+ \bigg( \frac{11}{36} - \frac{5}{72}\,g_N + \frac{53}{200}\,g_N^2\bigg)\ \,\Big(\frac{\mu}{m_N}\Big)^4
- \mu^2\,\Big(r^2_{E}+\frac{3}{4\,m_N^2}\Big)\,\Big(1-\frac{\mu^2}{m_N^2}\Big)\biggr] \,. \label{80}
\end{align}
\end{widetext}
It is worth mentioning that the spin-$0$ case can be obtained from
the spin-$1/2$ one by setting $g_N=0$ and redefining the charge radius.
We also note that the first two terms in powers of $\mu/m_N$ are universal and do not depend on the nuclear spin.

In addition to $E^{(4)}_\mathrm{fs}$ and $E^{(6)}_\mathrm{fs}$,
one needs to account for the one-loop electron vacuum polarization correction to the leading fine structure,
which can  be calculated as described in Ref.~\cite{muonicH}.
Our numerical results for the $2P$ fine structure of $\mu$He$^+$ are listed in Table \ref{fsmu}. They
are in agreement with the previous calculation of Karshenboim {\em et al}.~\cite{sgk2017, korzinin2018}
and with available experimental results \cite{pohl2016, schuhmann2023}.
The observed agreement supports the determination of the nuclear charge radii reported in these works.
This confirmation is important in view of a significant discrepancy in the
charge radii difference $r^2_E(h)-r^2_E(\alpha)$ between the electronic- and muonic-spectroscopy
determinations \cite{schuhmann2023, Kield2023, rmp}.

\section{Nuclear recoil in light muonic atoms} \label{sec viii}
In this section we examine the nuclear recoil correction for muonic atoms, as obtained within
two different approaches, namely, the leading-order $Z\alpha$ expansion result given by
Eqs.~(\ref{66}) and (\ref{68}), and the all-order (in $\Za$) approach. The comparison of results of the
two different methods will, first, validate the formulas derived in the present work and,
second, give us an idea about the higher-order (in $Z\alpha$) effects.

The general expression for the nuclear recoil correction in electronic and muonic atoms
valid to all orders in
$\Za$ was derived in Refs.~\cite{shabaev:85, pure_rec, shabaev:98:rectheo}. For a muonic atom,
it reads
\begin{align}\label{86}
E_{\rm rec} =&\, \frac{m_{\mu}^2}{m_N}\,\frac{i}{2\pi} \int_{-\infty}^{\infty}d\omega\,
\sum_n
 \frac1{\vare_a + \omega - \vare_n(1-i0)}
\nonumber \\ & \times
 \bra{a} \vec{p} - \vec{D}(\omega) \ket{n}
   \bra{n} \vec{p} - \vec{D}(\omega) \ket{a}\,,
\end{align}
where
$\vec{p}$ is the momentum operator,
$
D^j(\omega) = -4\pi Z\alpha \, \alpha^i \, D_C^{ij}(\omega,\vec{r})\,,
$
$\alpha^i$ are the Dirac matrices,
$D_C^{ij}$ is the transverse part of the photon propagator in the Coulomb gauge,
and the summation over $n$ is performed over the complete Dirac spectrum of a bound muon.
The photon propagator $D_C^{ij}$ describing the interaction between
a point-like and an extended-size particle was derived in Ref.~\cite{pachucki:23:prl}.

To separate out the contribution of order $\alpha^6$ and higher from $E_{\rm rec}$,
we subtract the contribution of previous orders. Specifically, we introduce the higher-order
remainder function $E_{\rm rec}^{(6+)}$, as follows
\begin{align}
E_{\rm rec}^{(6+)}  = &\ E_{\rm rec} - \frac{m_{\mu}^2}{m_N}
 \bigg[ \frac{(\Za)^2}{2n^2} + \frac{(\Za)^4}{2n^3}\Big( \frac1{j+\nicefrac12}-\frac1n\Big) \nonumber \\ &\
  + \frac{(\Za)^5}{\pi n^3}\,D_{50}\bigg]\,,
\end{align}
where $D_{50}$ is defined by Eq.~(\ref{10}),
\begin{align}
D_{50}(2p) = -\frac{8}{3}\,\ln [k_0(2p)]  - \frac{7}{18} = -0.308\,844\,332\ldots\,.
\end{align}

\begin{figure*}
\centerline{
\resizebox{0.9\textwidth}{!}{%
  \includegraphics{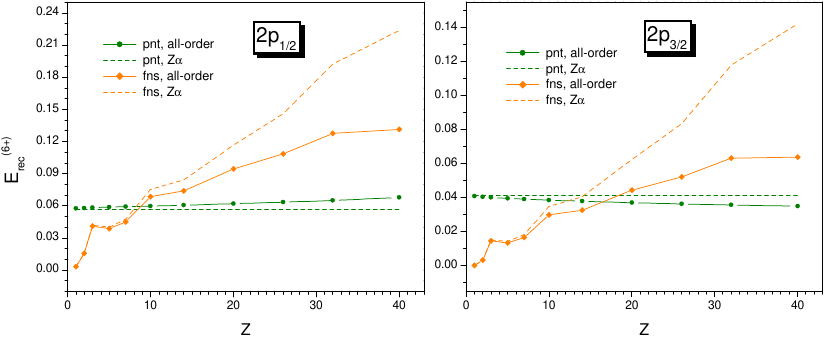}
}
}
\caption{
Nuclear-recoil point-nucleus and fns corrections $E^{(6+)}_{\rm rec}$
for the $2p_{1/2}$ and $2p_{3/2}$ states of muonic atoms,
as a function of the nuclear charge number $Z$. Units are $m_{\mu}^2/m_N\,(Z\alpha)^6$.
The nonsmoothness of the fns plots is due to the irregular dependence of the nuclear charge radius
on $Z$.
\label{fig:numrec}
}
\end{figure*}

We perform our numerical calculations of $E_{\rm rec}$ by the approach described in detail in
Ref.~\cite{yerokhin:23:rec}, for the exponential model of the nuclear charge
distribution. The total correction is conveniently separated into the point-nucleus (pnt) and
the finite-nuclear-size (fns) parts. The results
are presented in Table~\ref{tab:numrec} and Fig.~\ref{fig:numrec}. The numerical all-order
results are labeled as ``All-order'', whereas
the leading-order contributions obtained with Eqs.~(\ref{66}) and (\ref{68})
are labeled as ``$\Za$-exp''. We observe that the numerical
all-order results rapidly converge to the lowest-order analytical prediction
as $Z$ is decreased. The higher-order in $\Za$ corrections are quite small for the
point-nucleus contribution but become prominent for the fns correction already for
medium-$Z$ ions;
e.g., for $Z = 40$, the lowest-order fns formula overestimates the corresponding
all-order result by a factor
of about two.
It is also interesting that the fns part of $E_{\rm rec}$ rapidly grows with the nuclear charge
and dominates over the point-nucleus contribution for $Z>10$ for the $2p_{1/2}$ state
and $Z>20$ for the $2p_{3/2}$ state.

\begin{table*}
\caption{
Nuclear-recoil point-nucleus and fns corrections $E^{(6+)}_{\rm rec}$
for the $2p_{1/2}$ and $2p_{3/2}$ states of muonic atoms. Units are $m_{\mu}^2/m_N\,(Z\alpha)^6$.
 \label{tab:numrec}}
\begin{ruledtabular}
\begin{tabular}{llw{2.5}w{2.5}w{2.5}w{2.5}w{2.5}w{2.5}w{2.5}w{2.5}}
&&
                 \multicolumn{4}{c}{point}  &
                             \multicolumn{4}{c}{fns}  \\
\cline{3-6} \cline{7-10}
\\
  &&
                 \multicolumn{2}{c}{$2p_{1/2}$}  &
                             \multicolumn{2}{c}{$2p_{3/2}$}  &
                 \multicolumn{2}{c}{$2p_{1/2}$}  &
                             \multicolumn{2}{c}{$2p_{3/2}$}\\
\cline{3-4} \cline{5-6} \cline{7-8} \cline{9-10}
\\
\multicolumn{1}{c}{$Z$}
  &
       \multicolumn{1}{c}{$r_{E}$~[fm]}  &
                 \multicolumn{1}{c}{All-order}  &
                 \multicolumn{1}{c}{$Z\alpha$-exp.}  &
                 \multicolumn{1}{c}{All-order}  &
                 \multicolumn{1}{c}{$Z\alpha$-exp.}  &
                 \multicolumn{1}{c}{All-order}  &
                 \multicolumn{1}{c}{$Z\alpha$-exp.}  &
                 \multicolumn{1}{c}{All-order}  &
                 \multicolumn{1}{c}{$Z\alpha$-exp.} \\
\hline\\            %
   1  &0.8409&  0.05766   &     0.05729  & 0.04098   &     0.04167 &    -0.01046   &    -0.01057 &        -0.00094    &   -0.00107 \\
   2  &1.6755&  0.05801   &     0.05729  & 0.04060   &     0.04167 &    -0.05327   &    -0.05460 &        -0.01570    &   -0.01687 \\
   3  &2.4440&  0.05831   &     0.05729  & 0.04028   &     0.04167 &    -0.14986   &    -0.15665 &        -0.07078    &   -0.07637 \\
   5  &2.4060&  0.05882   &     0.05729  & 0.03971   &     0.04167 &    -0.13989   &    -0.14953 &        -0.06372    &   -0.07173 \\
   7  &2.5582&  0.05928   &     0.05729  & 0.03925   &     0.04167 &    -0.16372   &    -0.17963 &        -0.07837    &   -0.09168 \\
  10  &3.0055&  0.05992   &     0.05729  & 0.03866   &     0.04167 &    -0.25504   &    -0.29606 &        -0.13981    &   -0.17466 \\
  14  &3.1224&  0.06075   &     0.05729  & 0.03799   &     0.04167 &    -0.27241   &    -0.33449 &        -0.15007    &   -0.20346 \\
  20  &3.4776&  0.06205   &     0.05729  & 0.03715   &     0.04167 &    -0.34584   &    -0.47560 &        -0.19983    &   -0.31307 \\
  26  &3.7377&  0.06350   &     0.05729  & 0.03643   &     0.04167 &    -0.39010   &    -0.60553 &        -0.22868    &   -0.41778 \\
  32  &4.0742&  0.06519   &     0.05729  & 0.03580   &     0.04167 &    -0.44758   &    -0.81287 &        -0.26846    &   -0.58979 \\
  40  &4.2694&  0.06795   &     0.05729  & 0.03505   &     0.04167 &    -0.44082   &    -0.95617 &        -0.25970    &   -0.71121 \\
    \end{tabular}
\end{ruledtabular}
\end{table*}

\section{Summary}
We have derived the complete QED correction of order $\alpha^6$  to the binding energies of the $nP$  states of two-body systems consisting of the spin-$0$ or $1/2$ 
extended-size particles of arbitrary masses and magnetic moments. 
The derivation has been verified by an all-order in $\Za$ numerical calculation of the first-order in $m/M$ recoil contribution.
We have verified the  result for the positronium $l=1$ energies in Refs. \cite{Zatorski, Adkins, Czarnecki}
and verified previous calculations of the $2P$ fine splitting in light muonic atoms \cite{sgk2017, korzinin2018}.

The obtained formulas for the $l = 1$ states extend the previous $l>1$ results of Ref.~\cite{zatorski_22} and can be
applied to a wide class of two-body systems of immediate experimental interest, such as
hydrogen, hydrogen-like ions, muonic hydrogen, muonic helium ion, positronium,
muonium, {\em etc.}
In the future, even more exotic two-body atomic systems may become accessible for experimental studies, such as 
protonium and other hydrogen-like hadronic atoms \cite{venturelli2007}.
Comparisons of theoretical predictions of  these systems
in highly rotational states with accurate spectroscopic measurements would serve as tests of yet unexplored region of long-range interactions between hadronic particles.

The current theoretical predictions of energies of the $l>0$ levels of two-body systems
can be improved further by a calculation of the $\alpha^7$ correction, which is presently known in the nonrecoil limit only \cite{alpha7}, 
and by inclusion of the electron vacuum polarization in a nonperturbative manner as was done for muonic atoms \cite{muonicH}.

\acknowledgments
We are grateful to Jacek Zatorski for interesting discussions and comments, and to Clara Peset and Antonio Pineda for pointing out a mistake in our originally published results.

\appendix
\section{Matrix elements of various operators for $\bm P$-states}

Here we list results for 
matrix elements of various operators needed for our evaluation of $E^{(6)}$ for $nP$-states,
\begin{align}
\bigg\langle \frac{1}{r}\bigg\rangle = &\ \frac{\mu\,Z\alpha}{n^2}\,, \\
\bigg\langle \frac{1}{r^2}\bigg\rangle = &\ \frac{2\,(\mu\,Z\alpha)^2}{3\,n^3}\,,\\
\bigg\langle \frac{1}{r^3}\bigg\rangle = &\ \frac{(\mu\,Z\alpha)^3}{3\,n^3}\,,\\
\bigg\langle \frac{1}{r^4}\bigg\rangle = &\ 2\,(\mu\,Z\alpha)^4\bigg(\frac{1}{5\,n^3}-\frac{2}{15\,n^5}\bigg)\,,\\
\big\langle \,\vec p\;4\pi\,\delta^3(r)\,\vec p\,\big\rangle = &\ \frac{4\,(\mu\,Z\alpha)^5}{3}\,\bigg(\frac{1}{n^3}-\frac{1}{n^5}\bigg)\,,\\
\Big\langle\vec p\times 4\pi\,\delta^3(r)\,\vec p \Big\rangle = &\  i\, \frac{4\,(\mu\,Z\alpha)^5}{3} \bigg(\frac{1}{n^3}-\frac{1}{n^5}\bigg)\, \vec L \,,\\
\Big\langle \big(p^i 4\pi\,\delta^3(r) p^j\big)^{(2)}\! \Big\rangle = &  - \frac{4\, (\mu\,Z\alpha)^5}{3}\bigg(\frac{1}{n^3}-\frac{1}{n^5}\bigg)
\big(L^i L^j\big)^{(2)}\,.
\end{align}

\begin{widetext}
\section{Positronium $\bm P$-levels at the $\bm{\alpha^6}$ order}
\label{sec:app2}
The complete $\alpha^6$ correction to the energy levels of the $nP$-states of positronium is given by
\begin{align}
E^{(6)}_\mathrm{pos}(n^1P_1) =&\ m\,\alpha^6\left(-\frac{69 }{512\, n^6}+\frac{23 }{120\, n^5}-\frac{1}{12\, n^4}+\frac{163}{4320\, n^3}\right)\,,\\
	E^{(6)}_\mathrm{pos}(n^3P_0) =&\ m\,\alpha^6\bigg(-\frac{69}{512\, n^6}+\frac{119}{240\, n^5}-\frac{1}{3\, n^4}-\frac{833}{4320\, n^3} -\frac{a_1^2 + 6\,a_2}{24\,\pi^2\,n^3}\bigg)\,,\\
E^{(6)}_\mathrm{pos}(n^3P_1) =&\ m\,\alpha^6\bigg(-\frac{69}{512\, n^6}+\frac{77}{320\, n^5}-\frac{25}{192\, n^4}+\frac{553}{17\,280\, n^3}+\frac{a_1^2-2\,a_2}{48\,\pi^2\,n^3}\bigg)\,,\\
E^{(6)}_\mathrm{pos}(n^3P_2) =&\ m\,\alpha^6\bigg(-\frac{69}{512\, n^6}+\frac{559}{4800\, n^5}-\frac{169}{4800\, n^4}+\frac{17\,977}{432\,000\, n^3}+\frac{-a_1^2+18\,a_2}{240\,\pi^2\,n^3}\bigg) \,.
\end{align}
where $a_1$ and $a_2$ are the expansion coefficients of the electron magnetic-moment anomaly $a$, 
\begin{align}
a =&\ \frac{\alpha}{\pi}\,a_1 + \biggl(\frac{\alpha}{\pi}\biggr)^2\,a_2 +\ldots\,,\\
a_1 =&\ \frac{1}{2}\,,\\ 
a_2 =&\ \frac{3}{4}\,\zeta(3) -\frac{\pi^2}{2}\,\ln 2 +\frac{\pi^2}{12} + \frac{197}{144}\,.
\end{align}
The presented formulas agree with \cite[Eq.~(204)]{Zatorski} for all states. 
Note that in this section we switched to the literature definition of $E^{(6)}_\mathrm{pos}$ and included contributions from the
expansion of $g$-factors in $\alpha$, originating from $E^{(4)}$  in Eq.~(\ref{06}).

\end{widetext}
\end{document}